 \documentclass[conference]{IEEEtran}
 \IEEEoverridecommandlockouts
\usepackage{float}
\usepackage{amssymb,bm,cite,graphicx, fixmath, texdraw}
\usepackage{epsfig}
\usepackage{epstopdf}
\usepackage[english]{babel}
\usepackage{algorithm}
\usepackage{algorithmic}
\graphicspath{{./}{EPSFiles/}}
\usepackage{caption}
\usepackage{amsmath}

\usepackage{graphicx}
\usepackage{subfigure} 

\usepackage{amsmath}
 \makeatletter
\def\ifundefined{\@ifundefined}

\newcommand{\beq}{\begin{equation}}
\newcommand{\eeq}{\end{equation}}
\newcommand{\bea}{\begin{eqnarray}}
\newcommand{\eea}{\end{eqnarray}}

\newcommand{\nn}{\nonumber}
\newcommand{\nnl}{\nonumber \\}
\newcommand{\fig}[1]{Fig.\ \ref{#1}}

\DeclareMathOperator*{\argmin}{arg\,min}
\DeclareMathOperator*{\argmax}{arg\,max}

\def\bmat{\left[ \begin{array}}
\def\emat{\end{array} \right]}

\def\bmatt{\left\{ \begin{array}}
\def\ematt{\end{array} \right.}

\def\bset{\left\{ \begin{array}}
\def\eset{\end{array} \right\}}

\def\bpar{\left( \begin{array}}
\def\epar{\end{array} \right)}

\renewcommand{\theequation}{\arabic{equation}}
\renewcommand{\baselinestretch}{1}

\begin{document}

\title{A General Framework for Performance Analysis of Spatial Modulation over Correlated Fading Channels}
\author{\IEEEauthorblockN{Mutlu Koca}
\IEEEauthorblockA{Electrical $\&$ Electronics Engineering Dept. \\
Bo\v{g}azi\c{c}i University\\
 Bebek 34342 Istanbul, Turkey\\
Email: mutlu.koca@boun.edu.tr}
\and
\IEEEauthorblockN{Hikmet Sari}
\IEEEauthorblockA{SUPELEC\\
Plateau du Moulon - 3 rue Joliot-Curie\\
91192 Gif-sur-Yvette, France\\
E-mail:hikmet.sari@supelec.fr.}}

\maketitle

\begin{abstract}
We present a general method for the error analysis of spatial modulation (SM) systems over correlated and uncorrelated Rayleigh and Rician fading channels. The proposed method, making use of the properties of proper complex random variables and vectors, provides an exact upper bound for the class of fading channels considered for any number of  transmit and receive antennas and for a wide family of linear modulation alphabets. Theoretical derivations are validated via simulation results.
\end{abstract}

\section{Introduction}

Spatial modulation (SM) has emerged as a new and highly effective multiple-input multiple-output (MIMO) communication technology \cite{mes, jeg}. The fundamental principle of SM is simple: Both the signal and the  antenna spaces are used to convey information. In other words, at each transmission instant the incoming information bits are split into two groups. The first group is mapped onto a transmit antenna index whereas the second group is used to  select a constellation symbol within the modulation alphabet. Then only the antenna designated by the antenna bits is activated to send the selected symbol. This approach has a number of advantages: i) Because one antenna is activated at each time instant, not only the inter-channel interference (ICI) is avoided but also the receiver complexity is reduced. ii) Given a spectral efficiency of $R$ b/s/Hz.,  because SM allocates some bits for the antenna index instead of mapping all $R$ bits to the constellation symbols, the  constellation size is reduced in comparison to single antenna transmission. This increases the minimum distance between constellation points and results in an improved error performance, especially when $R$ is large. Notice that, space-shift keying (SSK) proposed in \cite{jeg2} as a special case of SM where information is transmitted only over the antenna space, (i.e., SM with a constellation size of $1$) have similar performance with even lower complexity, especially at low spectral efficiency. On the other hand, SSK is required to double the number of transmit antennas for every increment in the spectral efficiency. As a result SM is still a more viable alternative for practical number of antennas.

Despite its advantages,  the unique nature of SM systems,  that is, the transmission and reception of two different forms of information simultaneously, creates a challenge in their understanding and analysis. One challenging aspect is the performance analysis of SM under optimal detection. Originally  a performance analysis is presented for Rayleigh fading channels in  \cite{mes}, which is shown to be suboptimal in \cite{jeg}. Then in  \cite{jeg}, a maximum likelihood (ML) detection rule is presented along with a performance analysis for Rayleigh fading channels using the well-known union bound method of \cite{pro}.  It is also shown that  this upper bound can be computed with a closed form expression in the case of real constellations and uncorrelated fading conditions. However for complex constellations and over correlated fading channels,  it needs to be computed numerically using the Monte Carlo methods.  One alternative to numerical averaging methods is to employ the Chernoff or other exponential upper bounds as done in \cite{basar, basar2}, but these bounds are not tight especially in the case of correlated channels and higher spectral efficiencies. Notice that the performance analysis of SSK poses a relatively simpler problem compared to SM since only the antenna indices are transmitted as information. That is why, accurate closed form upper bounds are derived for Rayleigh fading channels in \cite{jeg2},  for correlated Rician channels in \cite{diren2}  and for single receiver Nakagami fading channels in \cite{diren3}. Because of its inherent difficulty  in the general MIMO setup,  the SM performance analysis problem is studied with only one receive antenna in  \cite{diren4} for uncorrelated and in \cite{diren} for correlated Rayleigh fading channels.  Among these, \cite{diren4} is particularly important as the authors make a detailed analysis of the error metric and show that by employing a Gaussianity approximation it is possible to derive tight upper bounds for certain digital modulation schemes such as phase-shift keying (PSK) and multilevel quadrature amplitude modulation (QAM). However, the analytical results of \cite{diren4} and \cite{diren} are not extendable to general SM schemes with multiple receive antennas and/or with other modulation schemes such as rectangular QAM.

To the best of our knowledge, performance analysis of SM  with an arbitrary number of transmit/receive antennas and general family of linear modulation alphabets is still an open problem. In the error analysis of conventional communication systems, the envelope of channel fading effects (in scalar or vector form) appears in the error performance metric. The distribution of this envelope or its square can be obtained for most known fading conditions. Then using probability density function (PDF), more commonly moment generating function (MGF) \cite{alo1, alo2} or characteristic function (CHF) \cite{anna} approaches it is possible to compute the average error rate for a wide class of digital modulation methods. As pointed out \cite{jeg, diren4, diren}, the difficulty in SM arises from the fact that the error metric consists of the envelope of not a single fading variable. Instead it contains a complex weighted mixture of two complex random vectors formed by the channel fading coefficients where the mixture weights are the constellation symbols.  Even in the uncorrelated Rayleigh fading case, if the constellation points  are not real, the real and imaginary parts of the resulting vector are not statistically independent. This makes it challenging to obtain a PDF or MGF (or CHF) for the envelope and therefore to derive a closed form expression for the average pairwise error probability (APEP).

In this paper, we propose an upper bounding technique for SM that is applicable to an arbitrary number of transmit/receive antennas and general modulation alphabets for uncorrelated/correlated Rayleigh and Rician fading channels. We extend the insight of \cite{jeg} that is limited to SM with real constellations to the case of complex constellations. Specifically, we show that for a number of fading scenarios, the random variable/vector that appears in the SM error metric  is a ``proper complex random variable/vector'' defined in \cite{nees}  and can be described with a joint multivariate Gaussian PDF. Then using its multivariate distribution instead of that of its envelope and employing the approach in  \cite{veer}, we present a framework to compute the upper bound for a class of fading channel models. We assume a general parametric correlated Rician channel model that can also be used to describe uncorrelated Rician and  Rayleigh fading conditions. The general framework can be utilized to derive error upper bound for a large class of channel fading conditions. The framework provides an exact bound generally via simple numerical integration and in some cases results in closed-form bounds. 

The rest of this paper is organized as follows. In Section II, the system and generalized channel fading model is presented. Then a brief review of proper complex random variables is given and the performance bounds are derived for MIMO SM  systems in Section III.  This paper is ended with  conclusive remarks and discussions in Section IV.


\section{System Model}

\subsection{Signal Model}
We consider the conventional SM system model with $N_t$ transmit and $N_r$ receive antennas employing optimal detection in \cite{jeg}. We also assume that the number of transmit antennas is an integer power of $2$, i.e., $N_t=2^n$, and the transmitter employs $M$-ary digital modulation to an $m$-bit message where $M=2^m$ and the modulation symbol set is  $\bm{{\mathcal X}}=\bset{ccccc} X_1, & \ldots & X_k, & \ldots & X_M \eset $.  Notice that the spectral efficiency of the system is $R=n+m=\log_2(N_tM)$. At each transmission instant,  each set of $n+m$ of bits is split into groups of $n$ and $m$ bits, and the former is used to select one of the $N_t$ antennas and the latter to be mapped onto one of $M$ possible complex constellation points determined by the particular digital modulation method. The method used in the bit-to-antenna index and bit-to-symbol mappings is immaterial to the central discussion, so throughout the paper we assume uniform mapping. The transmitted signal vector is ${\mathbf x}=\bmat{ccccc} x_1, & \ldots & x_\ell, & \ldots & x_{N_t} \emat^T $ where all but one entry is zero because only one antenna is active for transmission. Notice that if the $\ell$-th antenna is selected, then all entries other than $x_\ell$ is zero and $x_\ell \in \bm{{\mathcal X}}$. In other words, the position of the non-zero element denotes the antenna index and its value indicates the transmitted symbol. Similar to \cite{jeg}, we assume a power constraint of unity, i.e., $E_{\bf x}\left[{\bf x}^\dagger{\bf x}=1\right]$. 
The received signal model is expressed as
\beq
{\mathbf y}=\sqrt{\rho} {\mathbf H}{\mathbf x} + {\bm \nu} \label{rec_sig}
\eeq where  $\rho=E_s/N_0$ is the average signal to noise ratio (SNR) observed at each receiver branch, ${\mathbf y}$ and ${\bm \nu} $ are the $N_r \times 1$ received signal and channel noise vectors, respectively, and ${\bf H}$ is the $N_r \times N_t$ dimensional channel matrix. The elements of  ${\bm \nu}$ are modelled as independent identically distributed (i.i.d.) complex Gaussian variables with zero mean and unit variance, i.e., ${\nu}_k \sim{\mathcal{CN}}(0, 1)$ for $k=1,\ldots,N_r$.

\subsection{Fading MIMO Channel Model}
We assume a slow fading MIMO channel model with the sum of an average (or fixed, possibly line-of-sight) component and a variable (or random) component. Accordingly, $N_r \times N_t$ dimensional channel matrix ${\mathbf H}$ is described as
\beq
{\mathbf H} = \sqrt{\frac{K}{K+1}}\bar{{\bf H}} +\sqrt{\frac{1}{K+1}} \tilde{{\bf H}} \label{ch_model1}
\eeq where $\bar{{\bf H}}$ and  $ \tilde{{\bf H}}$  are the fixed and variable components, respectively. The square-root terms are the normalization weights with $K$ being the Rician factor and reflecting the ratio of the fixed and variable channel components. The fixed part $\bar{{\bf H}}$ is modelled as an all-one matrix, i.e., $[\bar{{\bf H}}]_{u,v}=\bar{h}_{u,v}=1$, for $u=1,\ldots,N_r$ and $v=1, \ldots, N_t$. The variable component of the channel matrix, $ \tilde{{\bf H}}$, consists of -possibly correlated- complex Gaussian variables.  Given $[\tilde{{\bf H}}]_{u,v}=\tilde{h}_{u,v}$, for all channel coefficient pairs $(\tilde{h}_{u,v}, \tilde{h}_{\hat{u}, \hat{v}})$ ($u, \hat{u}=1,\ldots,N_r$ and $v,\hat{v}=1, \ldots, N_t$) we assume that
\bea
\mbox{E}\left[ \tilde{h}^{R}_{u,v} \tilde{h}^R_{\hat{u},\hat{v}}\right]&=&\mbox{E}\left[ \tilde{h}^{I}_{u,v} \tilde{h}^I_{\hat{u},\hat{v}}\right],\nnl
\mbox{E}\left[ \tilde{h}^{R}_{u,v} \tilde{h}^I_{\hat{u},\hat{v}}\right]&=&\mbox{E}\left[ \tilde{h}^{I}_{u,v} \tilde{h}^R_{\hat{u},\hat{v}}\right]=0,\nn
\eea that is to say, the auto-correlations of the real and imaginary parts are the same and there is no correlation between real and imaginary parts. With this condition, the correlated channel matrix can be described by the well-known Kronecker correlation model in which  $ \tilde{{\bf H}}$ is expressed as
\beq
 \tilde{{\bf H}} = {\bf \Sigma}_r^{1/2} \breve{\mathbf H}{\bf \Sigma}_t^{T/2}  \label{kron}
\eeq where ${\bm{\Sigma}}_t$ and ${\bm{\Sigma}}_r$ are the real valued and Hermitian symmetric transmit and receive correlation matrices, respectively, with the elements defined as $[{\bm{\Sigma}}_t]_{u,\hat{u}}={\sigma}^t_{u,\hat{u}}$ for $u,\hat{u}=1,\ldots,N_t$, $[{\bm{\Sigma}}_r]_{v,\hat{v}}={\sigma}^r_{v,\hat{v}}$ for $v,\hat{v}=1,\ldots,N_r$. $\breve{\bf H}$ is the independent Rayleigh fading channel matrix with zero mean and unit variance. In other words, the elements of $\breve{\mathbf H}$ are described as independent identically distributed complex Gaussian random variables, i.e., $[\breve{\mathbf H}]_{u,v}=\breve{h}_{u,v}\sim{\mathcal{CN}}(0, 1)$ for $u=1,\ldots,N_r$ and $v=1, \ldots, N_t$. Notice that, with this channel model the multivariate conditional PDF of ${\bf y}$ can be written as:
\beq
 f_{\mathbf {Y}}({\bf y} \mid {\mathbf x}, {\mathbf H}) = \frac{1}{\pi^{N_t}\mbox{det}({\bf \Sigma})}\exp\big(-({\bf y}-{\bf m_Y})^\dagger{\bf \Sigma_Z}^{-1}({\bf y}-{\bf m_Y})\big) \label{ch_dist}
\eeq where ${\bf m_Y}=\mbox{E}\left[{\bf y}\right]=\sqrt{\frac{\rho K}{K+1}}\bar{{\bf H}}{\mathbf x}$ and ${\bf \Sigma}=\mbox{E}\left[({\bf y}-{\bf m_Y})({\bf y}-{\bf m_Y})^\dagger\right]=\frac{\rho}{K+1}{\bf \Sigma}_t\otimes{\bf \Sigma}_r$ are the mean vector and the covariance matrix, respectively (The operator $\otimes$ denotes the Kronecker product). Notice that the Kronecker correlation model separating the transmit and receive correlation effects has been used extensively within the context of SM/SSK literature \cite{diren2, basar, basar2}.  The correlation matrices can be formed according to a number of models as detailed in the discussion of the simulation parameters and results. 

Combining (\ref{ch_model1}) and (\ref{kron}), the general channel model can be rewritten as
 \beq 
{\mathbf H} = \sqrt{\frac{K}{K+1}}\bar{{\bf H}} +\sqrt{\frac{1}{K+1}}{\bf \Sigma}_r^{1/2} \breve{\mathbf H}{\bf \Sigma}_t^{T/2}  \label{ch_model}\eeq
which, via the appropriate selection of parameters, makes it possible to characterize  uncorrelated/correlated Rayleigh and Rician fading scenarios. 

\subsection{Optimal Detector and Average Probability of Error}

Given the conditional PDF in (\ref{ch_model}), the optimum antenna and symbol index pair  $(u_{\mbox{\small ML}}, v_{\mbox{\small ML}})$ in the maximum likelihood (ML) sense is expressed as
$$ (u_{\mbox{\small ML}}, v_{\mbox{\small ML}}) = \argmax_{u, v} \: \:  f_{\mathbf {Y}}({\bf y} \mid {\mathbf x}, {\mathbf H})  = \argmin_{u, v} \:\: D({\mathbf y}, {\mathbf h}_uX_v)
$$  where ${\mathbf h}_u$ is the $u$-th column of ${\mathbf H} $ for  $u=1,\dots, N_t$  and $X_v$ is the $v$-th element of the modulation alphabet $\bm{{\mathcal X}}$ for  $v=1,\dots, M$. $D({\mathbf y}, {\mathbf h}_uX_v)$ is the -modified- distance metric between ${\mathbf y}$ and ${\mathbf h}_uX_v$ defined as
\beq   D({\mathbf y}, {\mathbf h}_uX_v)= \sqrt{\rho}\parallel {\mathbf h}_uX_v \parallel^2-2\mbox{Re}\{ {\bf y}^\dagger {\mathbf h}_uX_v\}. \eeq
As shown in \cite{jeg}, the average bit error performance of the SM system with optimum detection can be computed by the well-known upper bounding technique in \cite{pro}. Accordingly, the average bit error probability (ABEP) is upper bounded as
\bea
\bar{P}_b \leq \frac{1}{N_tM-1} \sum_{u=1}^{N_t} \sum_{\hat{u}=1}^{N_t }\sum_{v=1}^M \sum_{\hat{v}=1}^M  \frac{N(u, \hat{u}, v, \hat{v})}{\log_2(N_tM)}  \bar{P}_s(u,\hat{u}, v, \hat{v}) \label{abep} \eea
where $N(u, \hat{u}, v, \hat{v})$ is the number of bits in error between the respective channel and symbol pairs, $({\mathbf h}_u,  X_v)$  and $({\mathbf h}_{\hat{u}}, X_{\hat{v}})$. The denominator term $\log_2(N_tM)$ in expression in (\ref{abep}) represents the total number of antenna and symbol bits and division with this term indicates the summation weight for the corresponding pairwise error probability (PEP). For large spectral efficiencies (bits per SM symbols) the average of the ratio $\frac{N(u, \hat{u}, v, \hat{v})}{\log_2(N_tM)}$ approaches to the factor of $\frac{1}{2}$, which is the one used in the derivations \cite{diren4}, \cite{diren}. In (\ref{abep}), $\bar{P}_s(u, \hat{u}, v, v)$ is the average pairwise symbol error probability (APEP) as  defined in equation (\ref{apep}).
\begin{figure*}[!htp]
 \bea \bar{P}_s(u, \hat{u}, v, \hat{v})  &=& {\large E}_{{\mathbf x}, {\mathbf H}}\left[ \mbox{Pr}\left\{D({\mathbf y}, {\mathbf h}_uX_v) > D({\mathbf y}, {\mathbf h}_{\hat{u}}X_{\hat{v}}) \mid  {\mathbf h}_u, {\mathbf h}_{\hat{u}}, X_v, X_{\hat{v}} \right\} \right] \nnl
&=&  {\large E}_{{\mathbf x}, {\mathbf H}} \left[\large{Q}\left(\sqrt{\parallel {\bf z} \parallel^2}\right)\right] \label{apep} \eea   \end{figure*}

In (\ref{apep}), the $N_r \times 1$ vector ${\mathbf z}$ is defined as \cite{jeg}:
\beq
{\mathbf z}=   \sqrt{\frac{\rho}{2}} \big({\mathbf h}_uX_v-{\mathbf h}_{\hat{u}}X_{\hat{v}}\big) \label{z}.
\eeq
Notice that the $k$-th element of ${\bf z}$, $k=1,\ldots,N_r$ can be defined as
\bea
z_k &=& \bar{z}_k+\tilde{z}_k =\sqrt{\frac{\rho}{2}} \big( h_{k,u} X_{v}- h_{k,\hat{u}}X_{\hat{v}}\big) \\
 \bar{z}_k &=& c\left[(X^R_v-X^R_{\hat{v}})+i(X^I_v-X^I_{\hat{v}})\right] \\
  \tilde{z}_k &=& c  \left(\tilde{h}^R_{k,u}X^R_v-\tilde{h}^I_{k,u}X^I_v-\tilde{h}^R_{k,\hat{u}}X^R_{\hat{v}}+\tilde{h}^I_{k,\hat{u}}X^I_{\hat{v}} \right) \nnl
 &+&\hspace{-.07in} i \:c \left(\tilde{h}^R_{k,u}X^I_v+\tilde{h}^I_{k,u}X^R_v-\tilde{h}^R_{k,\hat{u}}X^I_{\hat{v}}-\tilde{h}^I_{k,\hat{u}}X^R_{\hat{v}} \right)  \label{branch}
\eea where  the coefficient $c=\sqrt{\frac{\rho}{2(K+1)}}$, and $ \bar{z}_k$ and $\tilde{z}_k$ denote the fixed and variable components of $z_k$, respectively. 

Notice that in \cite{jeg}, it is shown that so long as the symbol constellation ${\bm{\mathcal X}}$ is real,  the real and imaginary parts of the expression in (\ref{branch}) are statistically independent of each other in the case of uncorrelated Rayleigh fading channels ($K=0$, $\tilde{\bf H}=\breve{\bf H}$). Then for any symbol pair $(X_v, X_{\hat{v}})$, the term $\parallel {\bf z} \parallel^2$ becomes a Chi-squared random variable with $2N_r$ degrees of freedom and therefore the average of the Q-function in (\ref{abep}) can be computed in closed form. On the other hand, as also indicated in \cite{jeg}, when $(X_v, X_{\hat{v}})$ come from complex constellations the distribution of the envelope of random variable $\tilde{z}_k$ (or $z_k$ in general) cannot be  easily obtained,  due to statistical dependency of the real and imaginary parts of $\tilde{z}_k$.  In this case, the Monte Carlo averaging methods are adopted in both \cite{jeg} and the subsequent works \cite{diren, diren4} for both SM and SSK error analysis. 

In the next section, we show that accurate statistical description of ${\bf z}$ is possible in terms of a joint multivariate PDF so long as it is a proper complex Gaussian random vector. Then using this distribution of ${\bf z}$ we provide a general framework in which the  APEP in (\ref{apep}) and the ABEP in (\ref{abep}) can be computed.

\section{Performance Analysis of Spatial Modulation}

\subsection{Proper Complex Random Vectors}


Let ${\bf Y}={\bf Y}^R+i{\bf Y}^I$ be a $N\times 1$ complex random vector with the mean ${\bf m_Y}= \mbox{E} \left[{\bf Y}\right]$. Suppose its real covariance matrices  ${\bm \Sigma}^R_{\bf Y} $, ${\bm \Sigma}^I_{\bf Y} $, ${\bm \Sigma}^{RI}_{\bf Y} $ and  ${\bm \Sigma}^{IR}_{\bf Y} $ and complex covariance and pseudo-covariance matrices ${\bm \Sigma_Y}$ and $\breve{\bm \Sigma_Y}$ are respectively given by the following equations as
\bea 
{\bm{\Sigma}}^R_{\bf Y} &=& \mbox{E} \left[({\bf Y}^R-{\bf m}^R_{\bf Y} )({\bf Y}^R-{\bf m}^R_{\bf Y} )^T \right]\nnl
{\bm{\Sigma}}^I_{\bf Y} & =& \mbox{E} \left[({\bf Y}^I-{\bf m}^I_{\bf Y} )({\bf Y}^I-{\bf m}^I_{\bf Y} )^T \right]\nnl
 {\bm{\Sigma}}^{RI}_{\bf Y}  &=& \mbox{E} \left[({\bf Y}^R-{\bf m}^R_{\bf Y} )({\bf Y}^I-{\bf m}^I_{\bf Y} )^T \right]\nnl 
 {\bm \Sigma}^{IR}_{\bf Y} &=& \mbox{E} \left[({\bf Y}^I-{\bf m}^I)_{\bf Y} ({\bf Y}^R-{\bf m}^R_{\bf Y} )^T \right]\nnl
{\bm{\Sigma}}_{\bf Y}  &=& \mbox{E} \left[({\bf Y}-{\bf m}_{\bf Y} )({\bf Y}-{\bf m}_{\bf Y} )^{\dagger} \right]\nnl
&=& ({\bm{\Sigma}}^R_{\bf Y}  +{\bm \Sigma}^I_{\bf Y})+i (  {\bm \Sigma_Y}^{IR} - {\bm \Sigma_Y}^{RI})\nnl
\breve{\bm{\Sigma}}_{\bf Y} &=& \mbox{E} \left[({\bf Y}-{\bf m}_{\bf Y} )({\bf Y}-{\bf m})^T_{\bf Y}  \right] \nnl
&=&({\bm \Sigma}^R_{\bf Y}  -{\bm \Sigma}^I_{\bf Y})+i ({\bm \Sigma}^{IR}_{\bf Y}  + {\bm \Sigma}^{RI}_{\bf Y}).\nn
\eea

{\bf Definition} \cite{nees}:\emph{ ${\bf Y}$ is said to be proper complex random vector if its pseudo-covariance matrix $\breve{\bm \Sigma_Y}={\bf 0}$, or equivalently, if ${\bm{\Sigma}}^R_{\bf Y}= {\bm{\Sigma}}^I_{\bf Y}$ and ${\bm{\Sigma}}^{RI}_{\bf Y}=-{\bm{\Sigma}}^{IR}_{\bf Y}$.}

A direct consequence of this condition is that \beq {\bm \Sigma}_{\bf Y} = 2{\bm \Sigma}^R_{\bf Y}+i 2{\bm \Sigma}^{IR}_{\bf Y}. \label{proper} \eeq

In the special case of ${\bf Y}$ being a scalar, denoted by $Y$, ${\bm \Sigma}^{IR}_Y={\bm \Sigma}^{RI}_Y$. If $Y$ is proper, which means $E\left[(Y-m_Y)^2\right]=0$, then ${\bm \Sigma}^{IR}_Y=-{\bm \Sigma}^{RI}_Y$ together with the condition above implies that ${\bm \Sigma}^{IR} = 0$, i.e., $Y^R$ and $Y^I$ are uncorrelated. This along with (\ref{proper}), implies that for a complex scalar $Y$ to be proper the real and imaginary components must have the same variance and be uncorrelated.  

Finally  if  ${\bf Y}$ is  a proper complex Gaussian vector, ${\bf Y}^R$ and ${\bf Y}^I$ are jointly Gaussian and  the joint multivariate PDF of ${\bf Y}$ is written as 
\beq  f_{\mathbf {Y}}({\bf y}) = \frac{1}{\pi^{N}\mbox{det}({\bm \Sigma}_{\bf Y})}e^{-({\bf y}-{\bf m_Y})^\dagger{\bm \Sigma}^{-1}_{\bf Y}({\bf y}-{\bf m_Y}) }.\label{y_dist}
\eeq 

Given the definitions of proper random vectors above and ${\mathbf z}$  in (\ref{z})-(\ref{branch}), it is easy  to show that ${\bf z}$ is a proper Gaussian random vector and derive the parameters of its PDF for both $N_r=1$ and $N_r>1$. 

When $N_r=1$, notice that the real and imaginary parts of $z$ are Gaussian random variables (real weighted sum of real Gaussian variables). Furthermore, 
\bea m_Z &=&  \sqrt{\frac{\rho K}{2(K+1)}}\left(X_v-X_{\hat{v}}\right) \nnl
{\bm \Sigma}_Z & =& 2{\bm \Sigma}^R_Z=2{\bm \Sigma}^I_Z= \sigma^2_z\nnl &=&\frac{\rho }{2(K+1)} \left[\mid X_v\mid^2+\mid X_{\hat{v}} \mid^2 -2\mbox{Re} \{\sigma^t_{u, \hat{v}}X_vX^*_{\hat{v}} \}\right]. \nn
\eea In other words,  the real and imaginary parts of $z$ are uncorrelated and the pseudo-variance of $z$ vanishes. This makes the variance real and therefore  $z$ a proper complex Gaussian random variable by definition. 

When $N_r>1$, notice from above that each element of the vector ${\bf z}$ consists of a proper complex Gaussian random variable with uncorrelated real and imaginary parts and identical -real- variances.  Given the receiver correlation matrix ${\bm\Sigma}_r$, and the definition of {\bf z} we can show that:
\bea 
{\bf m}_{\bf Z} &=&  \sqrt{\frac{\rho K}{2(K+1)}}\left(X_v-X_{\hat{v}}\right)  \label{mean_z} \\
 {\bm \Sigma}^R_{\bf Z} &=&{\bm \Sigma}^I_{\bf Z} \nnl
 &= &c^2\left[\mid X_v\mid^2+\mid X_{\hat{v}} \mid^2 -2\mbox{Re} \{\sigma^t_{u, \hat{v}}X_vX^*_{\hat{v}} \}\right]{\bm{\Sigma}}_r \nnl
 {\bm \Sigma}_{\bf Z} ^{IR}&=& {\bm \Sigma}_{\bf Z}^ {RI}={\bf 0} \nnl
 {\bm \Sigma}_{\bf Z} &=& 2 {\bm \Sigma}^R_{\bf Z} = 2 {\bm \Sigma}^I_{\bf Z}  \nnl
 &=& \hspace{-.1in}c^2 \left[\mid X_v\mid^2+\mid X_{\hat{v}} \mid^2 -2\mbox{Re} \{\sigma^t_{u, \hat{u}}X_vX^*_{\hat{v}} \}\right]{\bm{\Sigma}}_r \label{var_z} 
\eea

\begin{figure*}[!ht]
\bea
 \bar{P}_s(u, \hat{u}, v, \hat{v})&=& \frac{1}{\pi}\int_0^{\frac{\pi}{2}}\int_{\bf z}  \frac{1}{\pi^{N_r}\mbox{det}({\bm \Sigma}_{\bf Z})}   \exp\left(-\frac{{\bf z}^\dagger{\bf z}}{2\sin^2\theta}\right)\times \exp\left(-({\bf z}-{\bf m}_{\bf Z})^\dagger{\bm \Sigma}_{\bf Z}^{-1}({\bf z}-{\bf m}_{\bf Z}) \right) d{\bf z} d\theta \nnl
&=&  \frac{1}{\pi}\int_0^{\frac{\pi}{2}}\int_{\tilde{\bf{z}}}  \frac{1}{\pi^{N_r}\mbox{det}(\tilde{\bm \Sigma}_{\bf Z})}  \exp\left(-\frac{{\tilde{\bf{z}}}^\dagger{\tilde{\bf{z}}}}{\sin^2\theta}\right)  \exp\left(-({\tilde{\bf{z}}} -\tilde{\bf{m}}_{\bf Z})^\dagger\tilde{\bm{\Sigma}}_{\bf Z}^{-1}({\tilde{\bf{z}}} -\tilde{\bf{m}}_{\bf Z}) \right) d{\tilde{\bf{z}}}  d\theta. \nn \eea\end{figure*}

This  implies that the vector ${\bf z}$ is  a proper complex Gaussian vector with the conditional joint PDF given as
\beq  f_{\mathbf {Z}}({\bf z}\mid {\bf h}_u,{\bf h}_{\hat{u}}, X_v, X_{\hat{v}}) = \frac{e^{-({\bf z}-{\bf m}_{\bf Z})^\dagger{\bm \Sigma}_{\bf Z}^{-1}({\bf z}-{\bf m}_{\bf Z}) }}{\pi^{N_r}\mbox{det}({\bm \Sigma}_{\bf Z})} \label{z_dist}
\eeq where $\mbox{det}(\cdot)$ represents the matrix determinant operation and  ${\bf m}_{\bf Z}$ and ${\bm \Sigma}_{\bf Z}$ are as given in (\ref{mean_z}) and (\ref{var_z}), respectively.  Notice that proper complex random variables and vectors whose properties are presented in \cite{nees} do not require the real and imaginary components to be statistically independent. So long as the pseudo-covariance matrix vanishes, a proper complex random variable (vector) has a distribution.

\subsection{Upper Bound for the Error Performance of SM}

The distribution obtained in (\ref{z_dist}) can be used  to evaluate  the APEP in (\ref{apep}) by  following an approach similar to that of  \cite{veer}, such that
\bea \bar{P}_s(u, \hat{u}, v, \hat{v}) &=&\nnl
&&\hspace{-.6in}\int_{\mathbf z} \large{Q}\left(\sqrt{\parallel {\bf z} \parallel^2}\right) f_{\mathbf {Z}}({\bf z}\mid{\bf h}_u,{\bf h}_{\hat{u}}, X_v, X_{\hat{v}}) d{\bf z}. \label{pep}
\eea Then by using the properties of ${\bf z}$  and Craig's alternative definition of the $Q(\cdot)$ function, (\ref{pep}) is rewritten as in the integral formulations given the top of this page where the parameter change ${\tilde{\bf{z}}} \leftarrow {\bf z}/\sqrt{2}$ results in 
\bea \tilde{\bf m}_{\bf Z} &=& \frac{{\bf m}_{\bf Z}}{\sqrt{2}} = \sqrt{\frac{\rho K}{4(K+1)}}\big(X_v-X_{\hat{v}}\big),  \nnl 
\tilde{\bm \Sigma}_{\bf Z} &=& \frac{{\bm \Sigma}_{\bf Z}}{2}\nnl & =&  \frac{\rho }{4(K+1)} \left[\mid X_v\mid^2+\mid X_{\hat{v}} \mid^2 -2\mbox{Re} \{\sigma^t_{u, \hat{u}}X_vX^*_{\hat{v}} \}\right]{\bm{\Sigma}}_r. \nn\eea Notice as in  the second line of (\cite{nees}, eq. (6)), by the Hermitian symmetry property of $\tilde{\bm{\Sigma}}_Z$ and the fact that the joint PDF of a proper complex Gaussian vector integrates to unity, this integral then simplifies to the expression below
\beq  \bar{P}_s(u, \hat{u}, v, \hat{v}) =  \frac{1}{\pi}\int_0^{\frac{\pi}{2}}  
\frac{e^{-\tilde{\bf m}_{\bf Z}^\dagger[\tilde{\bm \Sigma}_{\bf Z}+\sin^2\theta{\bf I}]^{-1}\tilde{\bf m}_{\bf Z}} }{\mbox{det}\left(\frac{\tilde{\bm \Sigma}_{\bf Z}}{\sin^2\theta}+ {\bf I}\right)} d\theta. \label{pep2}
 \eeq Plugging this result in (\ref{abep}) gives the ABEP expression for SM employing any modulation alphabet.

The APEP integral  in (\ref{pep2}) simplifies to some closed form expressions in certain cases as we show in the sequel. Otherwise, especially in the general Rician fading case, i.e. when ${\bf m}_{\bf Z}\neq {\bf 0}$, no further simplification is possible but since (\ref{pep2}) is a single definite integral, it can be conveniently evaluated numerically for any given mean vector and covariance matrix without the need for any Monte Carlo averaging.  

\subsection{Special Case: Average Error Performance for SSK}

Notice from the definition of ${\bf z}$ in (\ref{z})   that under the special conditions that ${\bf h}_u={\bf h}_{\hat{u}}$ (same transmit antenna), $X_v=X_{\hat{v}}$ (same transmitted symbol), ${\mathbf z}$ remains to be a proper complex random vector. Of these, the latter is particularly important as it corresponds to the transmission of information bits with single symbol constellation, or equivalently, with SSK. Notice that in this case ${\bf z}$ becomes
\bea
{\mathbf z}&=& \sqrt{\frac{\rho}{2}} \big({\mathbf h}_uX_v-{\mathbf h}_{\hat{u}}X_{v}\big) = c \left(\tilde{\bf h}_{u}-\tilde{\bf h}_{\hat{u}}\right)X_{v} \label{SSK_z} \eea with the corresponding first and second order statistics
\bea 
{\bf m}_{\bf Z} &=&  {\bf 0} \nnl
{\bm \Sigma}_{\bf Z} &=&  \frac{\rho}{2(K+1)} \left(2- 2\mbox{Re}[\sigma^t_{u, \hat{u}}]\right){\bm{\Sigma}}_r. \label{SSK_st} \nn
\eea
The condition in (\ref{SSK_z}) implies that so long as the fixed components of the MIMO Rician channel are the same, their contributions  cancel out in the expression of  ${\bf z}$ and  the error performances over correlated Rayleigh and Ricean channels differ only with the scaling factor of $1/\sqrt{K+1}$. Accordingly, the ABEP for SSK can be written as 
\beq \bar{P}^{\mbox{SSK}}_b \leq\frac{1}{N_t-1} \sum_{u=1}^{N_t} \sum_{\hat{u}=1}^{N_t }\frac{N(u, \hat{u})}{\log_2(N_t)}  \bar{P}^{\mbox{SSK}}_s(u,\hat{u}) \nn \eeq where \beq  \bar{P}^{\mbox{SSK}}_s(u, \hat{u}) =  \frac{1}{\pi}\int_0^{\frac{\pi}{2}}  \left[\mbox{det}\left(\frac{\tilde{\bm \Sigma}^{\mbox{SSK}}_{\bf Z}}{\sin^2\theta}+ {\bf I}\right) \right]^{-1} d\theta \label{pep_SSK} \eeq with $\tilde{\bm \Sigma}^{\mbox{SSK}}_{\bf Z} =  \frac{\rho}{2(K+1)} \left(1- \mbox{Re}[\sigma^t_{u, \hat{u}}]\right){\bm{\Sigma}}_r.$  As a result the error upper bound derivations of SSK can be computed as that of  SM with M-PSK  with $M=1$.


\section{Simulation Results}

In this section, simulation results are presented to validate the proposed error performance calculation method for the SM/SSK systems under various channel fading conditions. For all simulations, a MIMO system with $4$ transmit and $4$ receive antennas is considered. We simulate transmission schemes with spectral efficiencies of $R=3$ b/s/Hz to $R=7$ b/s/Hz. In a $N_t=4$ system this corresponds to employing  modulation alphabets of $2$ up to $32$ symbols. For small $R$,  BSPK and QPSK constellations are used while for large $R$ values they are replaced by rectangular QAM constellations with $8$, $16$ and $32$ points. Notice that rectangular QAM constellations are chosen in order to show that the approach presented in this paper is not dependent on a particular constellation geometry. In all of the results presented, the curves corresponding to simulations are plotted with solid lines whereas those from analytical upper bound derivations are represented with dotted lines with identical markers. For reasons of clarity the marker of only one curve (simulation) is included in the legends. 

\begin{figure} \centering
\epsfxsize = 3.6 in \epsfysize = 3 in
\epsffile{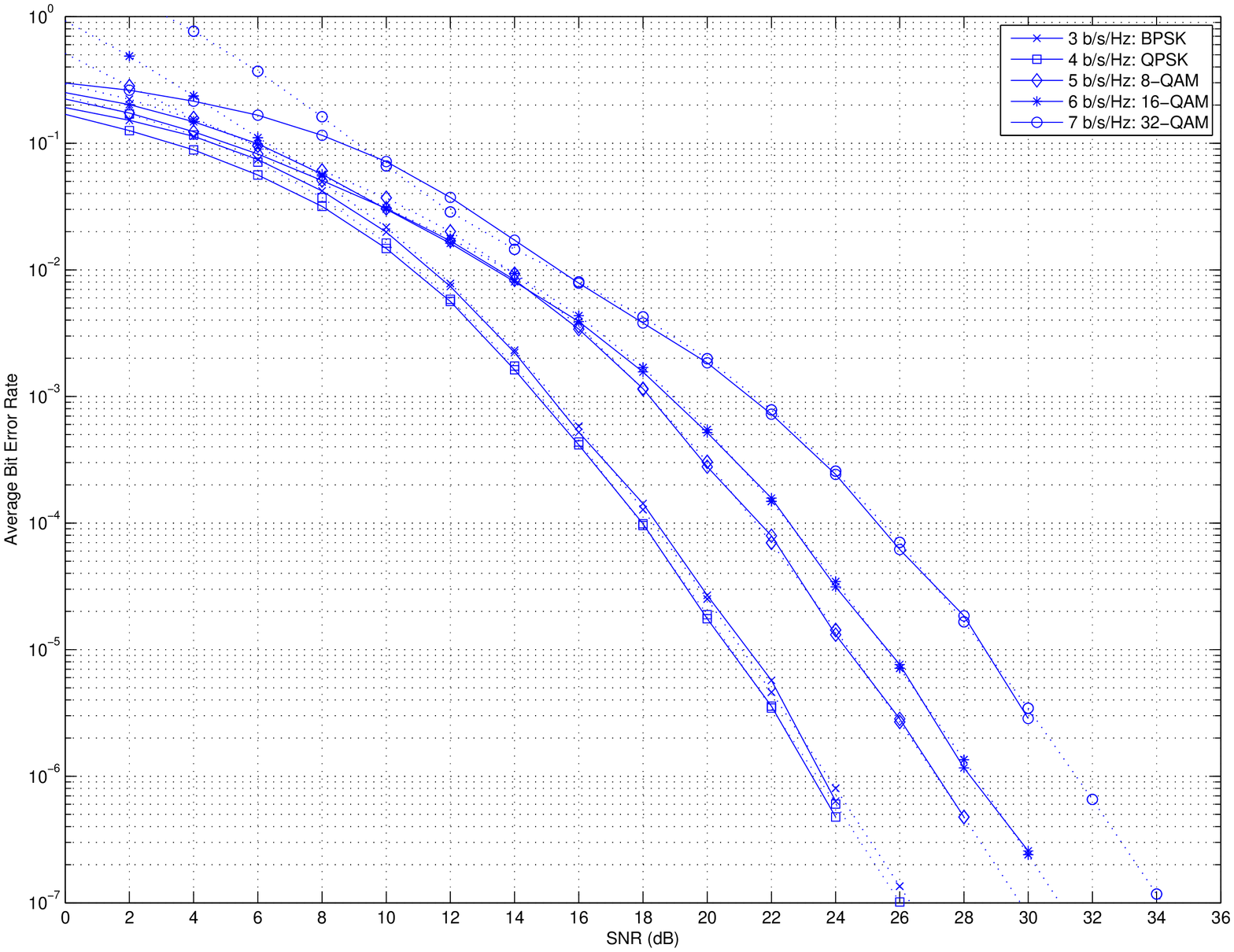}
\caption{SM error performance with $N_t=4$ and $N_r=4$ for$R=3$--$7$ b/s/Hz. Uncorrelated Rician fading with K=5.}  \label{fig:SM_4tx_4Nr_Ricean_K5}
\vspace{-0.1 in}
\end{figure}

\begin{figure} \centering
\epsfxsize = 3.6 in \epsfysize = 3 in
\epsffile{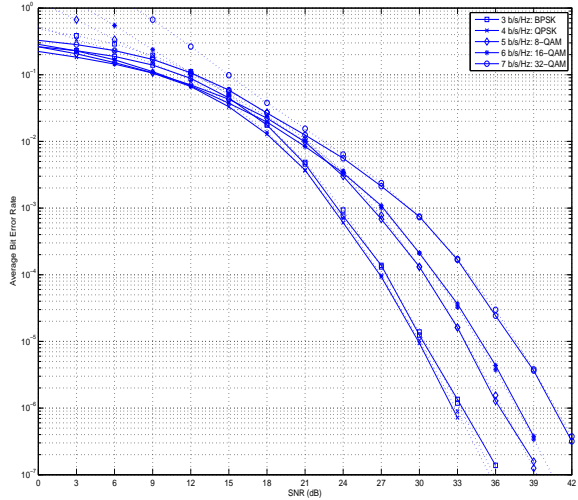}\caption{SM error performance with $N_t=4$ and $N_r=4$ for $R=3$--$7$ b/s/Hz. Correlated Rician fading, K=5. Exponential correlation model of \cite{loyka} with parameters $\gamma_t=\gamma_r=0.8$.}  \label{fig:SM_4tx_4Nr_Ricean_ecor_K5_gt08_gr08}
\end{figure}

\begin{figure}
\begin{center}
\epsfxsize = 3.6 in \epsfysize = 3 in
\epsffile{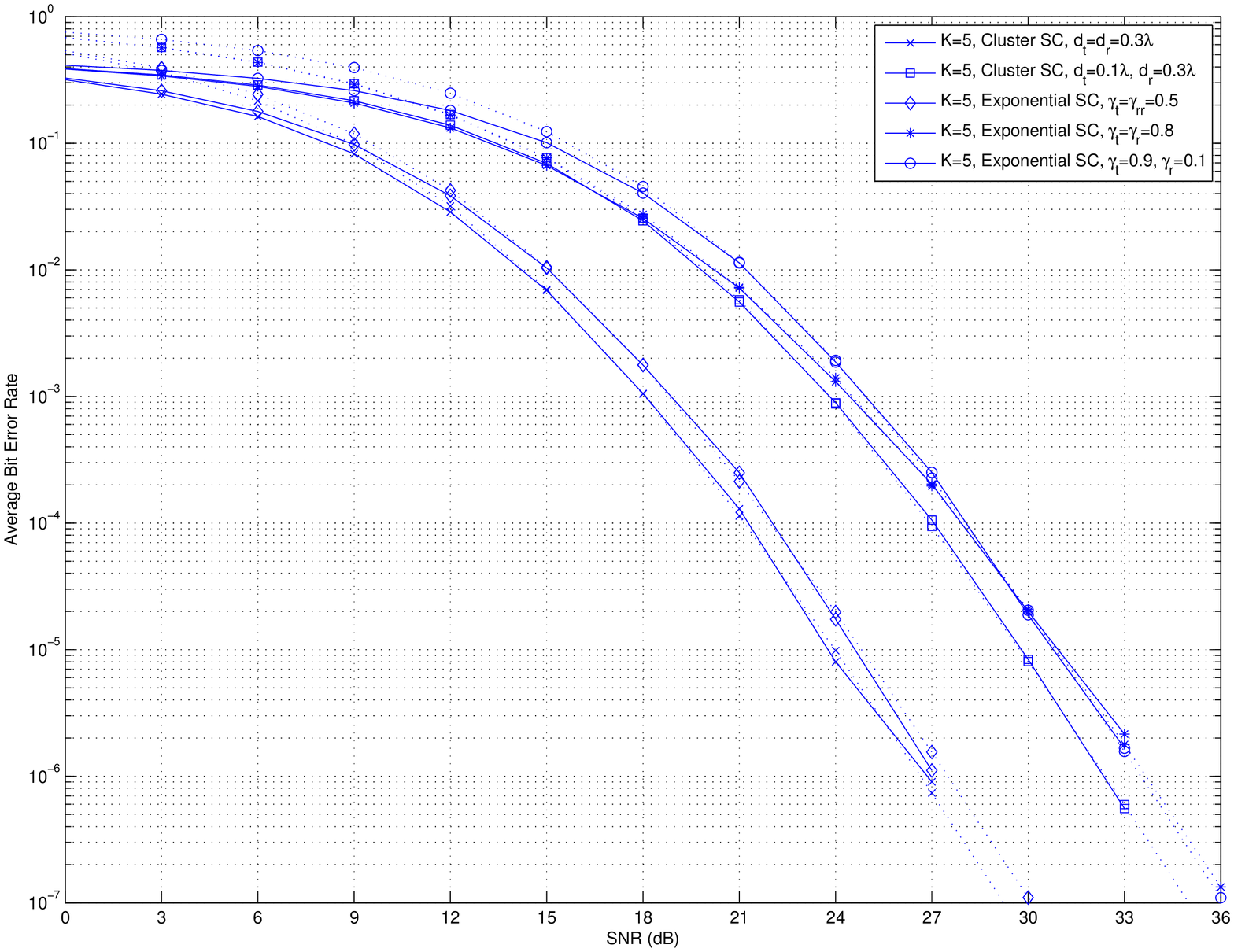}\end{center}
\caption{SSK error performance with $N_t=4$ and $N_r=4$ for correlated Rician fading, K=5.}  \label{fig:SSK_4tx_4Nr_Ricean_cor_K5}

\end{figure}

The transmit and receive channel correlation matrices in the Kronecker model are formed according to two common  and frequently used models. In the first, presented in \cite{foren}, the correlation matrices are computed based on a clustered channel model using the power azimuth spectrum distribution and the array geometry.  The exponential correlation model of \cite{loyka} is also considered where  the correlation matrix entries are formed as $ \left[{\bm{\Sigma}}\right]_{u,v}=\sigma_{u,v}= \gamma^{\mid u-v \mid} $ where $\beta$ is a fixed (real or complex) correlation coefficient between adjacent antennas. In simulations, we consider the cases where i) $\gamma_t=\gamma_r=0.5$, ii) $\gamma_t=\gamma_r=0.8$, and iii) $\gamma_t=0.9$,  $\gamma_r=0.1$. Notice that the first  and second cases are used to model moderate and strong  correlation on both sides, the third to describe heavy transmit correlation with almost no correlation at the receiver side.  

The theoretical error upper bound is compared with bit error performances obtained through numerical simulations in \fig{fig:SM_4tx_4Nr_Ricean_K5} -\fig{fig:SSK_4tx_4Nr_Ricean_cor_K5}. \fig{fig:SM_4tx_4Nr_Ricean_K5} shows performance of SM for uncorrelated Rician fading channels with a Rician factor of $K=5$ whereas the performance of SM under correlated Rician channels is shown in \fig{fig:SM_4tx_4Nr_Ricean_ecor_K5_gt08_gr08}.  Finally, \fig{fig:SSK_4tx_4Nr_Ricean_cor_K5} presents the SSK performance comparisons for $N_t=4$ ($R=2$ b/s/Hz spectral efficiency) under correlated Rician fading. Notice that in all SM simulations the results are depicted for the exponential correlation model \cite{loyka} where the clustered correlation model is omitted for space limitations. Both correlations models are depicted in the SSK simulations. All curves are drawn down to at least $1\times10^{-6}$ bit error rate (BER) levels and simulations indicate the upper bounds obtained with analytical framework are tight in all channel conditions considered. 

\section{Conclusion}

In this paper, we present a performance analysis framework for SM schemes over correlated Rayleigh and Rician fading channels characterized by  Hermitian symmetric real correlation matrices. Our framework is based on the observation that the complex vector used in the error function is proper complex Gaussian and has a multivariate joint distribution. We show that using this distribution  it is possible to obtain exact bounds on the performance of SM schemes and a number of special cases. One particular case is SSK, where the upper bound can be computed readily by considering it as SM with a constellation size of $1$. This upper bounding method for SSK is consistent with the physical realization of the transmission. Simulation results indicate a tight  match with the theoretical derivations.


\begin{thebibliography}{30}
\bibitem{mes} R. Mesleh, H. Haas, S. Sinanovic, C. W. Ahn and S. Yun, ``Spatial
modulation," {\it IEEE Trans. Veh. Technol.,} vol. 57, no. 4, pp. 2228--41, July 2008.
\bibitem{jeg} J. Jeganathan, A. Ghrayeb and L. Szczecinski,  ``Spatial modulation:
optimal detection and performance analysis," {\it IEEE Commun. Lett.,} vol. 12,
no. 8, pp. 545-547, Aug. 2008.
\bibitem{jeg2} J. Jeganathan, A. Ghrayeb and L. Szczecinski,  ``Space shift keying modulation for MIMO channels," {\it IEEE Trans. Wireless Commun.,} vol. 8, no. 3, pp. 3692-3703, Jul. 2009.
\bibitem{pro}  G. Proakis, {\it Digital Communications, 4th ed.,}  McGraw- Hill Higher Education, Dec. 2000.
\bibitem{basar} E. Basar, U. Aygolu, E. Panayirci, and H.V. Poor, ``Space-time block coded spatial modulation," {\it IEEE Trans. Commun.} vol. 59, no. 3,  pp. 823-832, Mar. 2011.
\bibitem{basar2} E. Basar, U. Aygolu, E. Panayirci, and H.V. Poor, ``New trellis code design for spatial modulation," {\it IEEE Trans. Wireless Commun.,} accepted for publication, Mar. 2011.
\bibitem{diren2} M. Di Renzo and H. Haas, ``Space shift keying (SSKÐ) MIMO over correlated Rician fading channels: Performance analysis and a new method for transmitÐdiversity'' {\it IEEE Trans. Commun.,}  vol.59, no. 1, pp. 2590--2603, Jan. 2011.
\bibitem{diren3} M. Di Renzo and H. Haas, ``A general framework for performance analysis of space shift keying (SSK) modulation for MISO correlated Nakagami-m fading channels,'' {\it IEEE Trans. Commun.,}  vol.58, no. 9, pp. 2590--2603, Sep 2010.
\bibitem{diren4} M. Di Renzo and H. Haas, ``Performance analysis of spatial modulation,'' in {\it Proc. CHINACOM,}  Aug. 25-27 2010, pp.1-7.
\bibitem{diren} M. Di Renzo and H. Haas, ``Performance comparison of different spatial modulation schemes in correlated fading channels,'' in {\it Proc. IEEE ICC 2010,}  May 2010, pp. 1Ð6.
\bibitem{alo1} M.-S. Alouini and A. Goldsmith, ``A unified approach for calculating error rates of linearly modulated signals over generalized fading channels,''  IEEE Trans. on Commun., vol. 47, no. 9, pp. 1324Ð1334, Sep 1999.
\bibitem{alo2} M. S. Alouini and M. K. Simon, {\it Digital Communications over Fading Channels, 2nd ed.}, John Wiley, New York, 2005.
\bibitem{anna} A. Annamalai, C. Tellambura, and V. K. Bhargava, ``A general method for calculating error probabilities over fading channels,Ó {\it IEEE Trans. Commun.,} vol. 53, no. 5, pp. 841Ð852, May 2005.
\bibitem{nees} F. D. Neeser and J. L. Massey, ``Proper complex random processes with applications to information theory,''  {\it IEEE Inform. Theory,}   vol. 39, no. 4, pp. 1293-1302, July 1993.
\bibitem{veer} V. Veeravalli, ``On performance analysis for signalling on correlated fading channels,"  {\it IEEE Trans. Commun.,}  vol.49, no. 11, pp. 1879-1883, Nov. 2001.
\bibitem{foren}  A. Forenza, D. J. Love,  and  R. W. Heath, Jr., ``Simplified Spatial Correlation Models for Clustered MIMO Channels with Different Array Configurations,'' {\it IEEE Trans. on Veh. Tech.,} vol. 56, no. 4, part 2, pp. 1924-1934, July 2007.
\bibitem{loyka} S. L. Loyka, ``Channel capacity of MIMO architecture using the exponential correlation matrix,Ó {\it IEEE Commun. Lett.,} vol. 5, pp. 369-371, Sept. 2001.
\end{thebibliography}
\end{document}